\documentclass[preprint,12pt]{elsarticle}

\usepackage{natbib}
\setcitestyle{authoryear,open={(},close={)}}
\usepackage{amssymb}
\usepackage{amsthm}
\usepackage{amsmath}
\usepackage{graphicx}
\usepackage[]{algorithm2e}
\usepackage{float}
\usepackage[a4paper]{geometry}
\usepackage[section]{placeins}
\usepackage{hyperref}
\usepackage{xcolor}
\usepackage{multirow}
\usepackage[flushleft]{threeparttable}
\usepackage{booktabs}
\usepackage{makecell}
\usepackage{placeins}

\newtheorem{theorem}{Theorem}
\newtheorem{definition}{Definition}

\newcounter{GameCtr}
\newenvironment{game}[1]{\vspace{2mm} \noindent\refstepcounter{GameCtr}\textbf{Game \theGameCtr} (#1) \ } {\hspace{\fill}$\spadesuit$}

\SetAlCapNameFnt{\footnotesize}
\SetAlCapFnt{\footnotesize}

\makeatletter
\def\ps@pprintTitle{%
  \let\@oddhead\@empty
  \let\@evenhead\@empty
  \let\@oddfoot\@empty
  \let\@evenfoot\@oddfoot
}
\makeatother

\begin{document}

\begin{frontmatter}



\title{Estimation of the Shapley value by ergodic sampling
\footnote{We thank P{\'e}ter Cs{\'o}ka, D{\'o}ra Gr{\'e}ta Petr{\'o}czy, Mikl{\'o}s Pint{\'e}r, Bal{\'a}zs Sziklai, and anonymous referees for helpful comments.}
}


 \author[1]{Ferenc Ill\'{e}s}
 \author[2]{P\'{e}ter Ker\'{e}nyi}

 \address[1]{Corvinus University of Budapest, ferenc.illes@uni-corvinus.hu, H-1093 Budapest, Fővám tér 8.}
 \address[2]{Corresponding author, Corvinus University of Budapest, peter.kerenyi@uni-corvinus.hu, H-1093 Budapest, Fővám tér 8.}


\begin{abstract}
The idea of approximating the Shapley value of an $n$-person game by Monte Carlo simulation was first suggested by \cite{mann1960values} and they also introduced four different heuristical methods to reduce the estimation error. Since 1960, several statistical methods have been developed to reduce the standard deviation of the estimate.
In this paper, we develop an algorithm that uses a pair of negatively correlated samples to reduce the variance of the estimate. Although the observations generated are not independent, the sample is \textit{ergodic} (obeys the strong law of large numbers), hence the name ``ergodic sampling". Unlike Shapley and Mann, we do not use heuristics, the algorithm uses a small sample to learn the best ergodic transformation for a given game. We illustrate the algorithm on eight games with different characteristics to test the performance and understand how the proposed algorithm works. The experiments show that this method has at least as low variance as an independent sample, and in five test games, it significantly improves the quality of the estimation, up to 75 percent.
\end{abstract}




\end{frontmatter}


\section{Introduction}
Since 1953 the Shapley value \citep{shapley1953value} is a fundamental solution concept of cooperative games with transferable utility (TU). As it is the sum of $2^{n-1}$ terms (where $n$ is the number of players), the exact value cannot be calculated in polynomial time of the number of players, based on its definition. In special classes of games, the Shapley value can be calculated in polynomial time (see for example \cite{dori1}, \cite{CostAlloc} and \cite{Airport}), but in general, it is intractable (see \cite{complex} or \cite{complex2}).

The Shapley value can be considered as the expected value of a discrete random variable. \cite{mann1960values} suggest estimating it with Monte-Carlo simulation. A consequence of the strong law of large numbers and the central limit theorem is that the mean of an independent sample is asymptotically normally distributed, unbiased and consistent estimate for the Shapley value. In practical applications, a good approximation can be almost as useful as the exact solution, and it is worth the effort to develop algorithms that provide approximations for the Shapley value with as low variance as possible. \cite{mann1960values} suggests four different heuristics to reduce the variance of the Shapley value, but only provides results for weighted voting games. \cite{Maleki} suggests reducing the estimation error by stratified sampling, and \cite{CASTRO2017} improves it with optimal allocation. \cite{touati2020bayesian} recommends a Bayesian Monte Carlo method for games with binary marginal contributions. It is also worth mentioning the deterministic multilinear approximation method of \cite{owen1972multilinear}, which is improved by \cite{leech2003}.

This paper proposes an algorithm that uses Monte Carlo simulation to estimate the Shapley value of any $n$-person game, if each marginal contribution can be calculated in polynomial time. In contrast to the algorithms using random independent samples, we use a ``strongly'' dependent but ergodic sample for the estimation. We generate samples in pairs that are highly negatively correlated. This sampling method is unbiased, consistent, and in the worst case, has asymptotically the same standard deviation as an independent sample with the same computational complexity, however, it can significantly reduce the variance of the sample mean compared to an independent sample.

The paper is organized as follows. In Section \ref{sec2}, we recall the necessary notations of the Shapley value and cooperative games in characteristic function form. In Section \ref{sec3} to \ref{sec5}, we describe a two-stage algorithm and its mathematical background. In the first stage, we use a greedy method to match the players to pairs, which yields an ergodic sample. In the second stage, we use this ergodic sample to estimate the Shapley value of the game. In Section \ref{sec6}, we provide computational results for eight test games, and analyze the results to understand the background of the mechanism, which leads to the improvements compared to the simple random sampling. Finally, in sections \ref{sec7}, we summarize our conclusions.

\section{Preliminaries}
\label{sec2}
A cooperative TU game in \textit{characteristic function} form is a pair $(N,v)$, where $N$ is a finite set of \textit{players} and the characteristic function $v:2^N\mapsto \mathbb{R}$ such that $v(\emptyset)=0$. The number of players is $|N|=n$. Subsets of $N$ are called coalitions. The characteristic function assigns the value to each coalition. The number of coalitions is $2^n$. 

The Shapley value assigns a unique value to each player as follows. Let $O_N$ be all possible \textit{orders} of the set of players $N$, i.e. \[O_N=\left\{o: N \longleftrightarrow \{1,2,\dots n\} \text{ bijective function }\right\}.\] 
If $o\in O_N$, $i \in N$ and $1\le j\le n$ then $o(i)=j$ means that player $i$ is the $j$th in order $o$. The number of all possible orders is, of course, $|O_N|=n! = |N|!$.
For each $o \in O_N$ and $i \in N$ let $Pre_o(i)$ be the set of players preceding player $i$ in order $o$ i.e.: \[Pre_o(i)=\{j \in N | o(j)<o(i)\}\,.\] 
The \textit{marginal contribution} of player $i$ in order $o$ is \[mc_o(i)=v(Pre_o(i) \cup \{i\})-v\left(Pre_o(i)\right)\,.\] The \textit{Shapley value} of player $i$ is the mean of the marginal contributions taking over all possible orders

\begin{equation}
\label{eq:Shapley}
Sh(i)=\frac{1}{n!}\sum \limits_{o \in O_N} mc_o(i)\,.
\end{equation}
It is easy to see that the Shapley value is an allocation of the value of the grand coalition: $\sum_{i \in N} Sh(i)=v(N)$. 

It is clear that any two terms in Equation \ref{eq:Shapley} are equal if the sets of players preceding player $i$ are the same: $mc_{o_1}(i)=mc_{o_2}(i)$ if $Pre_{o_1}(i)=Pre_{o_2}(i)$. If we make the summation over these equivalent terms, we obtain 
\[Sh(i)=\sum \limits_{S \subseteq N \setminus \{i\} } \frac{|S|!(n-|S|-1)!}{n!} \Big(v(S\cup \{i\})-v(S)  \Big)\,.\]

This formula reduces the number of terms from $n!$ to $2^{n-1}$, which does not help to calculate it when $n$ is large enough. However, by definition, the Shapley value is the expected value of a random variable over a finite (but large) probability space. This makes it possible to efficiently estimate it even in cases when the exact calculation is hopeless. 

In this paper, we address the following problem. Given an $n$-person game by its characteristic function $v$ that can be calculated in polynomial time, we wish to provide an approximation of the Shapley value of a given player $i \in N$.
We present an algorithm that uses a Monte Carlo method that is designed to reduce the variance of the sample mean. The main idea behind the algorithm is that it is not necessarily optimal to generate independent samples. Instead, correlated samples can have a lower variance.

\section{Theoretical background}
\label{sec3}

In this section, we recall the concept of ergodicity and prove a new theorem, which is the basis of our proposed method. 
\begin{definition}[Ergodicity]
	An infinite sequence of identically distributed random variables $X_1, X_2,\dots$ with finite expected value $\mu$ is said to be ergodic\footnote{Strictly speaking \textit{mean ergodic}, but we do not consider any other kind of ergodicity here.} if their sample mean converges to the expected value with probability 1:\[\mathbb{P}\left(\lim_{n\rightarrow \infty} \frac{ \sum X_i}{n}=\mu \right)=1\,.\]
\end{definition}

\noindent
Ergodicity means that we can estimate the expected value of the variable with the sample mean. By the strong law of large numbers, an independent sequence is always ergodic, but there are non-independent ergodic sequences as well.\\[5pt]
Now we consider the construction of an ergodic sequence. Let $\left(\Omega_1, \mathcal{A}_1, \mathbb{P}_1\right)$ be a classical probability space, that is, $|\Omega_1|<\infty, \ \mathcal{A}=2^{\Omega_1}$ and $\mathbb{P}_1(E)=\frac{|E|}{|\Omega_1|}$ \mbox{$ \ \forall E \subseteq \Omega_1$} and let \mbox{$X:\Omega_1\rightarrow \mathbb{R}$} be a random variable with $\mu = \mathbb{E}(X)$. \
Let $(\Omega,\mathcal{A},\mathbb{P})$ be the countable infinite power of $\Omega_1$ by the usual construction of the (infinite) product of measure spaces. For all $j \in \mathbb{N}$ and $\omega  = (\omega_1,\omega_2,\dots) \in \Omega$, let $X_j(\omega)=X(\omega_j)$. Observe that the variables $X_j$  are i.i.d. from the distribution of the original variable $X.$ Let \mbox{$t:\Omega_1 \mapsto \Omega_1$} be a bijective transformation and let \mbox{$T:\Omega \mapsto \Omega$} be the pointwise extension of $t$ to $\Omega$, i.e. $T(\omega)=T(\omega_1,\omega_2,\dots)=\left(t(\omega_1),t(\omega_2),\dots \right).$ Now let $K\in \mathbb{N}$ be a positive parameter and let us consider the following sequence of random variables
\begin{itemize}
	\item $\forall \ l \ge 1$, let $Y_{1,l} = X_l$
	\item for $2 \le k \le K$ and $l \ge 1$, let $Y_{k,l} = Y_{k-1,l}\circ T = Y_{1,l} \circ \underbrace{T \circ T \dots \circ T}_{k-1}=Y_{1,l} \circ T^{k-1}$
	\item Let the sequence $Y_j$ contain the variables $Y_{k,l}$ in the following (``column-wise'') order: \[Y_{1,1}, Y_{2,1},\allowbreak \dots Y_{K,1},Y_{1,2},Y_{2,2},\dots,Y_{K,2},Y_{1,3},Y_{2,3}, \dots \]
\end{itemize}

\noindent
For the rest of the paper, we will frequently switch between simple and double indices (whichever is more convenient) to denote the above sequence. Therefore by definition the symbols $Y_j$ and $Y_{l,i}$ stand for the same variable if $j = K\cdot (i-1) + l$ where $l\le K$. Our algorithm is based on the following observation:

\begin{theorem}
\label{tetel}
For each  integer $K \ge 2$ and bijective transformation $t:\Omega_1\longleftrightarrow\Omega_1$ the sequence $Y_j$ is ergodic.
\end{theorem}

\begin{proof}
\begin{sloppypar}
The measure of any cylinder set $C=(C_1,C_2,\dots) \in \mathcal{A}$ is \mbox{$\mathbb{P}\left(C\right)=\left(\frac{1}{|\Omega_1|}\right)^{|\{j|C_j\not = \Omega_1\}|}$}. As $t$ is surjective, $C_j=\Omega_1 \Leftrightarrow t(C_j)=\Omega_1$, therefore $\mathbb{P}\left(C\right)=  \mathbb{P}\left(T(C)\right)$. This means that $T$ is measure-preserving (it is enought to check this on the cylinders), which is the key of the proof. At first, we have to show that the variables are identically distributed\end{sloppypar}
\begin{multline*}
\mathbb{P}\left(Y_{k,l} \in B\right)=
\mathbb{P}\left(\{\omega \in \Omega | \omega \in Y^{-1}_{k,l}(B)\}\right)=
\mathbb{P}\left(\{\omega \in \Omega | T^{k-1}(\omega) \in Y^{-1}_{1,l}(B)\}\right)= \\=
\mathbb{P}\left(\{T^{k-1}(\omega) | T^{k-1}(\omega) \in Y^{-1}_{1,l}(B)\}\right) =
\mathbb{P}\left(\{\omega \in \Omega | \omega \in Y^{-1}_{1,l}(B)\}\right)=\\=
\mathbb{P}\left(Y_{1,l} \in B\right)=
\mathbb{P}\left(X_l \in B\right)=
\mathbb{P}\left(X_1 \in B\right)
\end{multline*}
for all $B\subseteq \mathbb{R}$ Borel-set. The third equation uses that $T$ is measure-preserving, the other equations are trivial. Now let $m \in \mathbb{N}$ be an integer such that $K|m$\footnote{Notation $K|m$ stands for $K$ is a divisor of $m$} and let $m=K\cdot L$.
\begin{multline*}
	\frac{\sum  Y_i}{m}=  \frac{Y_1+Y_{2}+\dots Y_{m}}{m}= \\
	\frac{(Y_{1,1}+Y_{2,1}+\dots +Y_{K,1})+(Y_{1,2}+\dots +Y_{K,2})+\dots+(Y_{1,L}+\dots +Y_{K,L})}{KL}=\\
	=\frac{\frac{\left(Y_{1,1}+Y_{1,2}+\dots +Y_{1,L}\right)}{L}+\frac{\left(Y_{2,1}+Y_{2,2}+\dots +Y_{2,L}\right)}{L}+\frac{\left(Y_{K,1}+Y_{K,2}+\dots +Y_{K,L}\right)}{L}}{K}
 =\frac{Z_1+Z_2+\dots+Z_K}{K}
\end{multline*}

\begin{sloppypar}
\noindent
It is enough to show that for $\forall \, 1 \le k \le K$ variables $Y_{k,1},Y_{k,2},\dots,Y_{k,L}$ are independent. In that case $\mathbb{P}\left(Z_k \rightarrow \mu\right)=1$, therefore $\mathbb{P}\left(\bigcap \limits_{k=1}^{K} \left\{\omega \in \Omega | Z_k(\omega) \rightarrow \mu \right\}\right)=1$. It is obvious that the mean of finitely many convergent series with the same limit is also convergent to the same number, so $\frac{\sum  Y_i}{m} \rightarrow \mu$. 
\end{sloppypar}

Note that the independence of variables $Y_{k,1},Y_{k,2},\dots,Y_{k,L}$ also implies that $\sqrt{L}(Z_j - \mu)$ is asymptotically normally distributed for all $j$ by the central limit theorem. The same argument does not hold for the mean of the $Z_js$ because they are not independent but negatively correlated. However, excluding degenerate cases, for example, when the variance of the sample mean is completely eliminated (it is possible), if there is any limit distribution for the sample mean, it can be nothing but normal. 

Variables $Y_{k,1},Y_{k,2},\dots,Y_{k,L}$ are by definition independent for $k=1$. For $k\ge 2$ this can be proved as follows:
\begin{multline*}
\mathbb{P}\left(\bigcap_{j=1}^{L} \left\{Y_{k,j} \in B_j\right\}\right)
=\mathbb{P}\left(\left\{\omega \in \Omega | \omega \in Y^{-1}_{k,1}(B_1), \omega \in Y^{-1}_{k,2}(B_2),\dots ,\omega \in Y^{-1}_{k,L}(B_L)\right\}\right)=\\
=\mathbb{P}\left(\left\{\omega \in \Omega | T^{k-1}(\omega) \in Y^{-1}_{1,1}(B_1), T^{k-1}(\omega) \in Y^{-1}_{1,2}(B_2),\dots,T^{k-1}(\omega) \in Y^{-1}_{k,L}(B_L)\right\}\right)=\\
=\mathbb{P}\left(\left\{T^{k-1}(\omega) | T^{k-1}(\omega) \in Y^{-1}_{1,1}(B_1), T^{k-1}(\omega) \in Y^{-1}_{1,2}(B_2),\dots,T^{k-1}(\omega) \in Y^{-1}_{k,L}(B_L)\right\}\right)=\\
=\mathbb{P}\left(\left\{\omega \ | \ \omega \in Y^{-1}_{1,1}(B_1), \omega \in Y^{-1}_{1,2}(B_2),\dots,\omega \in Y^{-1}_{k,L}(B_L)\right\}\right)=\\
=\prod_{j=1}^L \mathbb{P}\left(\left\{\omega \in \Omega \, | \, \omega \in Y^{-1}_{1,j}(B_j)\right\}\right)
=\prod_{j=1}^L \mathbb{P}\left(Y_{1,j} \in B_j\right)
=\prod_{j=1}^L \mathbb{P}\left(Y_{k,j} \in B_j\right)\,.
\end{multline*}
For all $B_1, B_2, \dots B_L \subseteq \mathbb{R}$ Borel-sets. Here the third line comes from that $T$ is measure-preserving, and the last equation holds because we have shown that all the variables are identically distributed. 

Now assume that $K \not | m$, and let $s=s(m)= \max\{j \le m \ | \  K | j\}$.

\begin{multline*}
 \frac{Y_1+Y_{2}+\dots Y_{m}}{m}= \frac{Y_1+Y_{2}+\dots Y_{s}}{m}+\frac{Y_{s+1}+\dots+Y_{m}}{m}=\\ =
\underbrace{\frac{Y_1+Y_{2}+\dots Y_{s}}{s}}_{\begin{tabular}{c}$\downarrow$ \\ $\mu$ \end{tabular}}\cdot\underbrace{\frac{s}{m}}_{\begin{tabular}{c}$\downarrow$ \\ 1 \end{tabular}}+\underbrace{\frac{Y_{s+1}+\dots+Y_{m}}{m}}_{\begin{tabular}{c}$\downarrow$ \\ 0 \end{tabular}}
\end{multline*}
As $K$ is a constant and $m-s \le K$ if $m \rightarrow \infty$ then so does $s$, so the first term goes to $\mu$ and $\frac{s}{m} \rightarrow 1$. The last term on the right side is bounded by $\frac{K \cdot \max{|X|}}{m} \rightarrow 0$. 
\end{proof}
\section{Two-stage algorithm}
\label{sec:4}
\begin{sloppypar}
Though the statement of Theorem \ref{tetel} is very abstract, it is easy to generate a sample from the variables $Y_j$. In our case $\Omega_1=O_N$, so we need an efficiently computable transformation $t:O_N\mapsto O_N$ that assigns orders of players to orders of players, and a parameter $K$. If these are given we can generate a sample as follows:
\begin{itemize}
	\item Generate a random order of players $o_{1,1}$ uniformly over $O_N$.
	\item Calculate $Y_{1,1} = mc_{o_{1,1}}(i)$
	\item Compute the new orders $o_{k,1}=t(o_{k-1,1})$ for $k = 2,\dots,K$
	\item Calculate $Y_{k,1} = mc_{o_{k,1}}(i)$
	\item Start again by generating a new order of players $o_{1,2}$, calculate $o_{k,2}$ and  $Y_{k,2}$ for each $k=2,\dots,K$ the same way.
	\item Calculate another $K$ elements $Y_{1,3}, Y_{2,3}, \dots, Y_{K,3}$, and so on.
\end{itemize}
We summarize this method in algorithm ErgodicShapley (see Algorithm \ref{alg:ergodic}), which generates a sequence of ergodic samples and uses it to estimate the Shapley value of any game. 
\end{sloppypar}
\begin{algorithm}
\label{alg:ergodic}
\caption{The pseudo-code of ErgodicShapley}
\SetKwFor{For}{For}{}{Next}
\DontPrintSemicolon

\hrule
\BlankLine
\TitleOfAlgo{ErgodicShapley($K$,$m$,$i$)}

\textbf{Input:} parameter $K \in \mathbb{N}$, \ size of the sample $m \in \mathbb{N}$ such that $m=KL$, player $i \in N$ 

\BlankLine
\hrule
\BlankLine

\For{$j:1 \ \KwTo \ L$}{
	 Generate a random order of players $o_{1,j}$\;
	\For {$k: 2 \ \KwTo \ K$}{
 		Calculate recursively new orders $o_{k,j}=t(o_{k-1,j})$.
	}
}
\BlankLine
We have $o_{1,1},o_{2,1},\dots,o_{K,1},o_{1,2},o_{2,2},\dots,o_{K,2},\dots,o_{1,L}\dots o_{K,L}$ non-independently generated orders. \;
\BlankLine

\For {$k:1 \ \KwTo \ K$}{ 
	\For {$l: 1 \ \KwTo \ L$}{
		 Let $Y_{k,l} = mc_{o_{k,l}}(i)$\
	}
}
$\hat{Sh}(i):=\frac{\sum \limits_{k,l} Y_{k,l}}{m}$ \;
\BlankLine
\Return $\hat{Sh}(i)$

\end{algorithm}

As a first example, let us consider the following game:\newpage

\begin{game}{non-symmetric voting game}
\label{game1}
\begin{sloppypar}
	Let $n\in \mathbb{N}$, $N=\{1,2,\dots,n\}$, \mbox{$w=(w_1,w_2,\dots,w_n)\in {\mathbb{R}}^n$} and $W \in \mathbb{R}$. For $S\subseteq N$ let \[v(S) = \begin{cases}1 & \text{if} \ \ \sum \limits_{i \in S} w_i > W,\\ 0 & \text{if} \ \ \sum \limits_{i \in S} w_i \le W. \end{cases}\]
	
The parameters of the game are set as follows: $n=51$, $w = (45,41,27,26,26,25,21,\allowbreak17,17, 14,13,13,\allowbreak  12,12,12,\allowbreak 11,\allowbreak  \underbrace{10,\dots,10}_{4},\allowbreak \underbrace{9,\dots,9}_{4},\allowbreak 8,8,\allowbreak \underbrace{7,\dots ,7}_{4},\allowbreak \underbrace{6,\dots,6}_{4},5, \allowbreak \underbrace{4,\dots,4}_{9}, \underbrace{3, \dots,3}_{7})$ and $W=\frac{\sum w_i}{2}= 269$ (this is the United States Electoral College in the '70s).
\end{sloppypar}

\end{game}

We still need a parameter $K \in \mathbb{N}$ and a transformation $t$ to run the algorithm. Later we will address the problem of finding them, for now, we ``guess'' an appropriate transformation $t$ to demonstrate the algorithm. Let $K=3$ and let $o\in O_N$ be an order of players and construct the following new order:

\[\forall i \in \{1,2,\dots,n=51\} \ \ t(o)(i)= \begin{cases}o(i)+17 & \text{if }\ o(i) \le 34 \\ o(i)-34& \text{if }\ o(i) > 34 \end{cases}\]

\noindent
We summarize this special case in five steps in Algorithm ShapleyK3.

\begin{algorithm}
\label{alg:K3}
\caption{ Special algorithm for the non-symmetric voting game  (Game \ref{game1}).}
\DontPrintSemicolon
\hrule
\BlankLine
\TitleOfAlgo{ShapleyK3 for non-symmetric voting game (Game \ref{game1})}
\textbf{Input:} $m \in \mathbb{N}$ size of the sample such that $3|m$, $i \in N$ player
\BlankLine
\hrule
\BlankLine

1. Generate a random order of players: $o_1$ \;
\BlankLine
2. Consider the arrival order of the players and construct the following new order: the last 17 people (last third of the players) arrive first in the same relative order, and the first 34 players go to the end of the row in the same relative order. Let this order be $o_2$.\;
\BlankLine
3. Construct order $o_3$ from $o_2$ in the same way.\;
\BlankLine
4. Repeat steps 1-3 $L = \frac{m}{3}$ times independently to get a sample size $m$.\;
\BlankLine
5. Calculate $cm_o(i)$ for all the orders and let $\hat{Sh}(i)$ be the sample mean. \;
\hspace{4mm} \Return $\hat{Sh}(i)$
\BlankLine
\BlankLine
\end{algorithm}

Table \ref{tab1} shows the results of the calculation of the Shapley value of the first player of the voting game. We can say that the standard deviation of the sample means is reduced by about 10\%. This reduction is not impressive, but it is easy to see that this is the best possible result using three correlated samples\footnote{In general if $X$ and $Y$ have same Bernoulli distribution with $E(X) =E(Y) = p< 1/2$, the minimum correlation between them occurs when E(XY) = 0 and it is $-p/(1-p)$. In this case, the expected value is the Shapley value of the first player, which is $p \approx 0.088$. If we calculate the covariance matrix of three such variables, we can see that the standard deviation of their mean is about $89\%$ compared to the independent variables.}

\begin{table}[H]
\begin{center}
{\renewcommand{\arraystretch}{1.3}
	\begin{tabular}{ l  c c }
		\hline
		Sample size & $m=10^5-1$ & $m=10^6-1$ \\ 
		\hline
		Standard deviation of independent samples &  0.0008568 & 0.0002890 \\
		Standard deviation of the new algorithm & 0.0008012 & 0.0002609 \\
		Ratio & 0.9351519 & 0.9027682 \\
		\hline
	\end{tabular}
}
\end{center}
\caption{The table shows the results of algorithm ShapleyK3 (Algorithm \ref{alg:K3}). Standard deviation of the ergodic sample mean is reduced by about 10\% compared to an independent sample in case of Game \ref{game1}.}
\label{tab1}
\end{table}

Note that the proposed ergodic algorithm in the present form is not efficient: The sum of the estimated Shapley values does not necessarily equal the grand coalition's value if we use independent samples for each player. We can make the ergodic estimation efficient, as \cite{CASTRO2017} suggested in the case of the stratified sampling.
Our ergodic sampling method uses a $K$-length block of samples to estimate the Shapley value. The first block of samples can be the same for all players, which can be considered as simple random sampling, from which we can obtain $Sh_1^s, Sh_2^s,\dots Sh_n^s$ estimates for the Shapley values, which is efficient. This requires practically no extra computation, but only one $K$th part of the sample is used. After that, we can generate the rest of the samples, so we can calculate the Shapley value estimation $Sh_1^{erg}, Sh_2^{erg},\dots Sh_n^{erg}$ from the ergodic sample, which has lower variance. The formula 
\[Sh_i^{eff} = Sh_i^{erg} + \left(v(N) - \sum_j Sh_j^{erg}\right)\frac{Sh_i^{s}}{v(N)}\]
combines these two estimates into an efficient one.

\section{Algorithm to find a suitable transformation}
\label{sec5}
\begin{sloppypar}
The algorithm is not useful if we do not have an intuition on how to define transformation $t$. Our goal in this section is to provide a general algorithm to construct it for a given game. The algorithm presented in Section \ref{sec:4} generates variates this way: \[\underbrace{Y_{1,1},Y_{2,1},\dots,Y_{K,1}},\underbrace{Y_{1,2},Y_{2,2}\dots,Y_{K,2}},\dots,\underbrace{Y_{1,L},Y_{2,L}\dots,Y_{K,L}}\]
A $K$-length block is constucted by generating the first element randomly, then compute the other $K-1$ ones based on the transformation, then repeat it $L$ times. Assuming that variates in the different blocks are independent, the variance of the mean of the full sequence is \[\sigma^2 = \frac{L \sum \limits_{1\le k_i,k_j \le K} C_{k_i,k_j} }{m^2}\,,\] where $C$ is the covariance matrix of variables \[Y_{1,1}, \ \ Y_{2,1} ,\ \  Y_{3,1}, \ \ \dots \ \ ,Y_{K,1}\,.\]
So our task is to find a transformation such that these variables are as negatively correlated as possible. In general, determining the optimal K and transformation seems a challenging problem. Although particular well-functioning heuristics can be developed for each game class utilizing its unique characteristics (as we have done before for the non-symmetric voting game) that significantly improve the algorithm's efficiency, in this paper, we now present a ``silly" but general transformation finding procedure that works for each game class. Our goal in this paper is primarily to test the theoretical procedure based on the ergodic sampling, so we choose this not necessarily effective procedure, which works in all game classes. Finding optimal transformations could be the topic of later research and practitioners' tasks to develop heuristics that work well for their practical problems. 

Now we address the special case of K=2. Algorithm GreedyK2 tries to find \mbox{$\left(o_1, o_2\right)\in O_N \times O_N$} pairs such that $mc_{o_1}(i)$ and $mc_{o_2}(i)$ have minimal empirical correlation. Let \mbox{$S_n=\{\pi:\{1,2,\dots,n\}\longleftrightarrow\{1,2,\dots,n\} \text{ bijection } \}$} be the \mbox{$n$-order} symmetric group. It is clear that for any $o \in O_N$ and $\pi \in S_n$ $(\pi \circ o) \in O_N$ i.e. for any fix $\pi \in S_n$ transformation $t_\pi:O_N\mapsto O_N$ for which $t_\pi(o)=\pi \circ o$ is a suitable transformation for \mbox{Theorem \ref{tetel}}. 

The algorithm constructs a permutation $\pi$  such that $\pi^K=\pi^2=id$, the identity of $S_n$. For each $a, b \in \{1,2,\dots,n\}$ let $\langle a,b\rangle \in S_n$ be the transposition of $a$ and $b$, i.e. the unique permutation that swaps $a$ and $b$ and the rest of the elements are mapped to themselves. If $o_1 \in O_N$ is an order of the players then $o_2=\langle a,b\rangle \circ o_1 \in O_N$ is a valid notation, and it means that we construct a new order by swaping the positions of the $a$th and $b$th player in order $o_1$.
In a nutshell, Algorithm GreedyK2 constructs permutation $\pi$ and transformation $t_\pi$ as follows.
\end{sloppypar}

\begin{itemize}
\item Choose a suitable parameter $m_1 \in \mathbb{N}$.
\item Generate orders of players $o_1,o_2,\dots o_{m_1}$ independently (with replacement). Let $\mathcal{O}$ be the set of these orders.
\item For each $a < b \in \{1,2,\dots,n\}$ let $\mathcal{O}_{a,b}= \left\{\langle a,b\rangle \circ o_j | o_j \in \mathcal{O}\right\}$.
\item For each $a < b \in \{1,2,\dots,n\}$ let $c_{a,b}$ be the empirical covariance between samples $\{mc_o(i)|o \in \mathcal{O}\}$ and $\{mc_o(i)|o \in \mathcal{O}_{a,b}\}$
\item Take the covariances in ascending order: $c_{a_1,b_1}\le c_{a_1,b_2}\le \dots\le c_{a_n,b_n}$. The length of this sequence is $\binom{n}{2}$.
\item Construct the first pair $(a_1,b_1)$ and delete $c_{a_1,b_1}$ from the list. 
\item Let $(a_2,b_2)$, $(a_3,b_3)$ etc. be always the first pair in the sequence that's members are both different from the previously chosen ones. Let $E_{op}$ be a list of these chosen pairs.
\item Let $\pi$ be the permutation that swaps the members of these pairs, i.e. $\pi = \prod \langle a_i, b_i \rangle$ and let $t_\pi$ be the following transformation: For each $\ o \in O_N$ let $t_\pi(o) = \underbrace{\left(\prod \limits_{(i,j) \in E_{op}} \langle i,j \rangle\right)}_{\pi} \circ o = \pi \circ o$ 
\end{itemize}

In other words, we construct a full graph over the set $\{1,2,\dots,n\}$ and give a weight for each edge by calculating an empirical covariance. Then we try to find a minimum weight matching by a greedy algorithm.\footnote{The Hungarian method could find an exact minimum weight matching, but that has higher time complexity and does not improve our algorithm.} Bullets six and seven describe the greedy algorithm: we take the edges in ascending order by their weights and always choose the next independent edge. The last four points of the algorithm can be rephrased as follows. Let $C$ be a matrix of the covariances calculated in the first four steps, containing the variance of the sequence $mc_o(i)$ in the diagonal. Let $a_1$ and $b_1$ be the row and the column of the minimum of the matrix. Record the pair $(a_1, b_1)$ and delete rows and columns $a_1$ and $b_1$ from the matrix (both rows and columns, because the matrix is symmetric). Now apply the same step to the smaller matrix until the matrix has no rows and columns to get a sequence of pairs $(a_j,b_j)$. Define $t_\pi$ as the transformation that swaps numbers $a_j$ and $b_j$.

Consider the number of steps the algorithm makes. It is easy to see that its magnitude is approximately at most $ m_1 \binom{n}{2}G_n\approx \frac{m_1n^2}{2}G_n$ where $G_n$ is the complexity of the game, i.e., the evaluation of the characteristic function $v$, if it is at least linear. Note that swapping the $a$th and $b$th player only changes player $i$'s marginal contribution if the swap constructs a new coalition, i.e., if $a \le o(i) \le b$ or vice versa. If both $a$ and $b$ are less or both are greater than $o(i)$, then $mc_o(i)$ does not change, so it is unnecessary to evaluate the characteristic function again. The $mc$ value changes only with probability $\frac{1}{3}$, so we can still divide the running time by a factor of $3$.

Combining the algorithm ErgodicShapley with the optimization algorithm GreedyK2, we obtain the final algorithm OptK2Shapley, which we analyze in the following sections. It uses $2m_2$ sample size. In the third step, $m_2$ is calculated such that the running time of the algorithm (including the greedy optimization) has the same time complexity as the simple random sampling with sample size $m$. 

\begin{figure}
	\begin{center}
		\includegraphics[scale = 0.75]{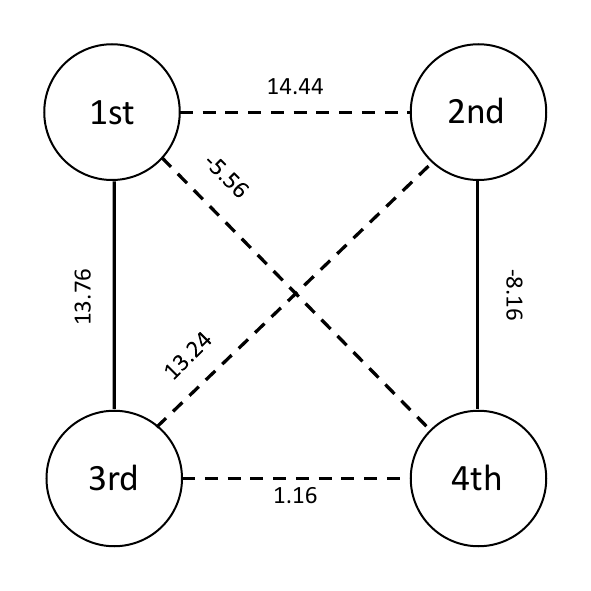}
		\caption{Weighted graph for Algorithm GreedyK2. Edges of the matching (solid lines) indicate that the algorithm suggest to swap the second and the fourth player as well as the first and third ones to generate correlated samples.}
		\label{K2Graph}
	\end{center}
\end{figure}

Let us illustrate the algorithm for a four-player game in which the value of a three-member coalition is 2, the value of the grand coalition is 12, and all other coalitions are worth 0. The players are called A, B, C, and D. We would like to estimate the Shapley value of player A. To 
find a suitable transformation, the algorithm randomly generates $m_1=10$ orders of players. The first column in \mbox{Table \ref{K2table}} shows these orders, and the second column shows the corresponding marginal contributions of player A. The algorithm then generates a new $m_1$ size set of orders by swapping the third and fourth players in the original order. The third and fourth columns of the table show these orders and the corresponding marginal contributions. Similarly, the algorithm swaps the second and fourth, the second and third, the first and fourth, and so on, players in the original order and creates ${n}\choose{2}$ new orders. For the sake of simplicity, we do not indicate all of these orders in \mbox{Table \ref{K2table}}, but the corresponding marginal contributions are shown in the fifth, sixth, and so on ninth columns. For example, we get the fourth element of column $mc_{2,3}$ if we swap the second and third players in order BACD (the fourth original order), which is BCAD, and calculate the marginal contribution of player A to the coalition \{B,C\}. When all these data are available, the algorithm creates a graph, where the set of nodes is $\{1,2,\dots,n\}$, and the weights on the edges are the covariances between columns $mc_0$ and $mc_{i,j}$ in Table \ref{K2table}. This graph is shown in Figure \ref{K2Graph}. Next, the algorithm greedily finds a minimum weight matching in this graph by first choosing pair $(2,4)$ and then $(1,3)$. This means that the transformation the algorithm will use is $t(1) = 3$, $t(2)  = 4$, $t(3) = 1$, and $t(4) = 2$. This is the output of GreedyK2. The algorithm generates $m_2$ brand new random orders, applies the transformation to each one, and uses these correlated samples to estimate the Shapley value. For example, a new order is ACDB. The marginal contribution of player A in this order is 0. Now swap the second player with the fourth one and the first one with the third one. We get order DBAC, where the marginal contribution of player A is $2$. It is worth pointing out that this transformation always turns the marginal contribution to zero if it is positive and turns it to nonzero if it is zero, therefore we expect a negative correlation between the samples. We have indicated in the last column of Table \ref{K2table} the marginal contributions we obtain if we apply this transformation to the original orders in the first column. The correlation between columns $mc_0$ and $mc_t$ is $-53\%$, so the algorithm performs well in this example. The estimation of the Shapley value of player A is $2.6$ based on the first column and $3.4$ based on the last one. If we average them, we incidentally obtain the exact value of 3.

\begin{table}
\centering
\begin{tabular}{l r l r r r r r r r}
\hline
$\text{Order}_0$ & $mc_0$ & $\text{Order}_{3,4}$ & $mc_{3,4}$ & $mc_{2,4}$ & $mc_{2,3}$ & $mc_{1,2}$ & $mc_{1,3}$& $mc_{1,4}$ & $mc_t$\\
\hline
CBAD & 2  \ \ & CBDA & 10 \ \ & 2 \ \  & 0 \ \ & 2 \ \ & 0 \ \ & 2 \ \ & 0 \ \ \\
DABC & 0 \ \ & DACB &   0 \ \ & 10 \ \ & 2 \ \ & 0 \ \ & 0 \ \ & 0 \ \ & 10 \ \ \\
BDAC & 2 \ \ & BDCA & 10 \ \ & 2 \ \  & 0 \ \ & 2 \ \  & 0 \ \ & 2 \ \ & 0 \ \ \\
BACD & 0 \ \ & BADC &   0 \ \ & 10 \ \  & 2 \ \ & 0 \ \ & 0 \ \ & 0 \ \ & 10 \ \ \\
ABDC & 0 \ \ & ABCD &   0 \ \ & 0 \ \  & 0 \ \ & 0 \ \ & 2 \ \ & 10 \ \ & 2 \ \ \\
BADC & 0 \ \ & BACD &   0 \ \ & 10 \ \ & 2 \ \ & 0 \ \ & 0 \ \ & 0 \ \ & 10 \ \ \\
DCBA &10 \ \ & DCAB &  0 \ \ & 0 \ \  & 10 \ \ & 10 \ \ & 10 \ \ & 0 \ \ & 0 \ \ \\
BCAD & 2 \ \ & BCDA & 10 \ \ & 2 \ \  & 0 \ \ & 2 \ \ & 0 \ \ & 2 \ \ & 0 \ \ \\
ABCD & 0 \ \ & ABDC &   0 \ \ & 0 \ \  & 0 \ \ & 0 \ \ & 2 \ \ & 10 \ \ & 2 \ \ \\
DBCA &10 \ \ & DBAC &  2 \ \ & 0 \ \  & 10 \ \ & 10 \ \ & 10 \ \ & 0 \ \ & 0 \ \ \\
\hline
\end{tabular}
\caption{Illustration of Algorithm GreedyK2. Column $mc_0$ contains the marginal contributions of player A in permutations in column 1. Column $mc_{j,l}$ contains marginal contributions of permutations if we swap players in position $j$ and $l$. Column $mc_t$ is the marginal contributions of permutations after swapping all pairs found by the algorithm.}
\label{K2table}
\end{table}

\FloatBarrier

\begin{algorithm}
\caption{The pseudo-code of GreedyK2.}
\label{alg:greedy}
\DontPrintSemicolon
\SetKwFor{For}{For}{}{Next}
\SetKwFor{While}{While}{Do}{{End While}}

\hrule
\BlankLine
\TitleOfAlgo{Algorithm GreedyK2}
\textbf{Input:} a positive parameter $m_1 \in \mathbb{N}$, player $i\in N$  \;
\BlankLine
\hrule
\BlankLine
Let $G=(V,E)$ be an undirected full graph, where the set of nodes is $V=\{1,2,\dots,n\}$ the set of edges is $E=V\times V$, and let $\beta_{r,s} \ \  (r,s \in V)$ be weights on the edges (do not have to be initialized).\;
Generate independently and uniformly $o_1,o_2\dots,o_{m_1}$ random orders of players.

\For{$j:1 \ \KwTo \ m_1$}{

			Calculate $X_j = mc_{o_j}(i)$
		}

\For {$r:1 \ \KwTo \ n$} {
	
	\For{$s:r \ \KwTo \ n$}{

		\For{$j:1 \ \KwTo \ m_1$}{

			Let $o^{(r,s)}_j = \langle r,s \rangle \cdot o_j$ \;
			Calculate $Y^{(r,s)}_j = mc_{o^{(r,s)}_j}(i)$
		}
		
		 Let $\beta_{r,s}=\beta_{s,r}= Cov(X,Y^{(r,s)})$ \ (covariance of data sets $X$ and $Y^{(r,s)}$)		

	}
}

Let $E_{op}$ be an empty list of edges\;
Take the edges $E$ in increasing order by the $\beta$ weights. In case of a tie, loop edges are preferred.\;
\While{$E_{op}$ is not a matching}{
	Append to $E_{op}$ a minimum weight edge $e_i$ such that it is not adjacent to any previously choosen $e_j \ (j < i)$ edges. 
}
For any $o \in O_N$ let $t(o)(i) = j \Leftrightarrow (i,j) \in E_{op}$\;
\textbf{Return} $t$\;
\BlankLine
\BlankLine
\end{algorithm}

\begin{algorithm}
\caption{Algorithm OptK2Shapley.}
\label{alg:optShapley}
\DontPrintSemicolon
\hrule
\BlankLine
\TitleOfAlgo{Algorithm OptK2Shapley}
\textbf{Input:} sample size $m \in \mathbb{N}$, player $i\in N$\;
\hrule
\BlankLine
Choose a parameter $m_1$ and run algorithm \textit{GreedyK2} to find $t$.\;
Generate $m_2 = \frac{m-m_1\frac{n^2}{6}}{2}$ random orders independently.\;
For each order $o$ generated in the previous step calculate $t(o)$\;
For all the $2m_2$ orders $o$ calculate $mc_o(i)$.\;
Take the mean of the generated $2m_2$ numbers.\;
\BlankLine
\BlankLine
\end{algorithm}

\section{Results}
\label{sec6}
We expect a variable performance of the algorithm depending on the different characteristics of the games. We used similar games frequently employed in the literature (\cite{CASTRO2009}, \cite{Maleki}), plus three other specific games to compare the results to the natural benchmark, simple independent random sampling.

\subsection{Test games}

For the analysis, we use the following test games (plus Game \ref{game1} introduced in Section \ref{sec:4}):

\begin{game}{symmetric voting game}
\label{game2}
This is a special case of the non-symmetric voting game (Game \ref{game1}) introduced in Section \ref{sec:4}. Each weight is set to $w_i=\frac{1}{n}$ where  the number of players is $n=100$.
\end{game}

\begin{game}{shoes game}
\label{game3}
Let the set of players $N=\{1,2,\dots,100\}$, characteristic function $v(S)=\min(S\cap N_{left},S\cap N_{right}) \ \forall S\subseteq N$ where $N_{left} =\{1,2,\dots,50\}$, $N_{right} =\{51,52,\dots,100\}$.
\end{game}

\begin{game}{airport game}
\label{game4}
Let $N=\{1,2,\dots,100\}$ and $c= (\underbrace{1,\dots,1}_{8},\underbrace{2,\dots,2}_{12},\allowbreak \underbrace{3,\dots ,3}_{6},\underbrace{4,\dots,4}_{14},\allowbreak\underbrace{5,\dots,5}_{8},\underbrace{6, \dots,6}_{9},\underbrace{7, \dots,7}_{13},\underbrace{8, \dots,8}_{10},\underbrace{9, \dots,9}_{10},\underbrace{10, \dots,10}_{10})$. The characteristic function of the game is $v(S)=\max\{c_i | i \in S\}$.
\end{game}

\begin{game}{minimum spanning tree game}
\label{game5}
Let $G$ be the weighted graph on \mbox{Figure \ref{minimumspanningtreerajz}}. The set of nodes is $V=\{0,1,2,\dots,100\}$, the edges are defined as follows: \[E= \left\{(x,x+1) | x\in \{1,2,\dots ,99\right\}\} \cup \{(100,1)\} \cup \bigcup_{i \in \{1,2,\dots,100\}}\{(0,i)\} \subseteq V \times V.\]

\begin{figure}[H]
	\begin{center}
		\includegraphics[scale=0.5]{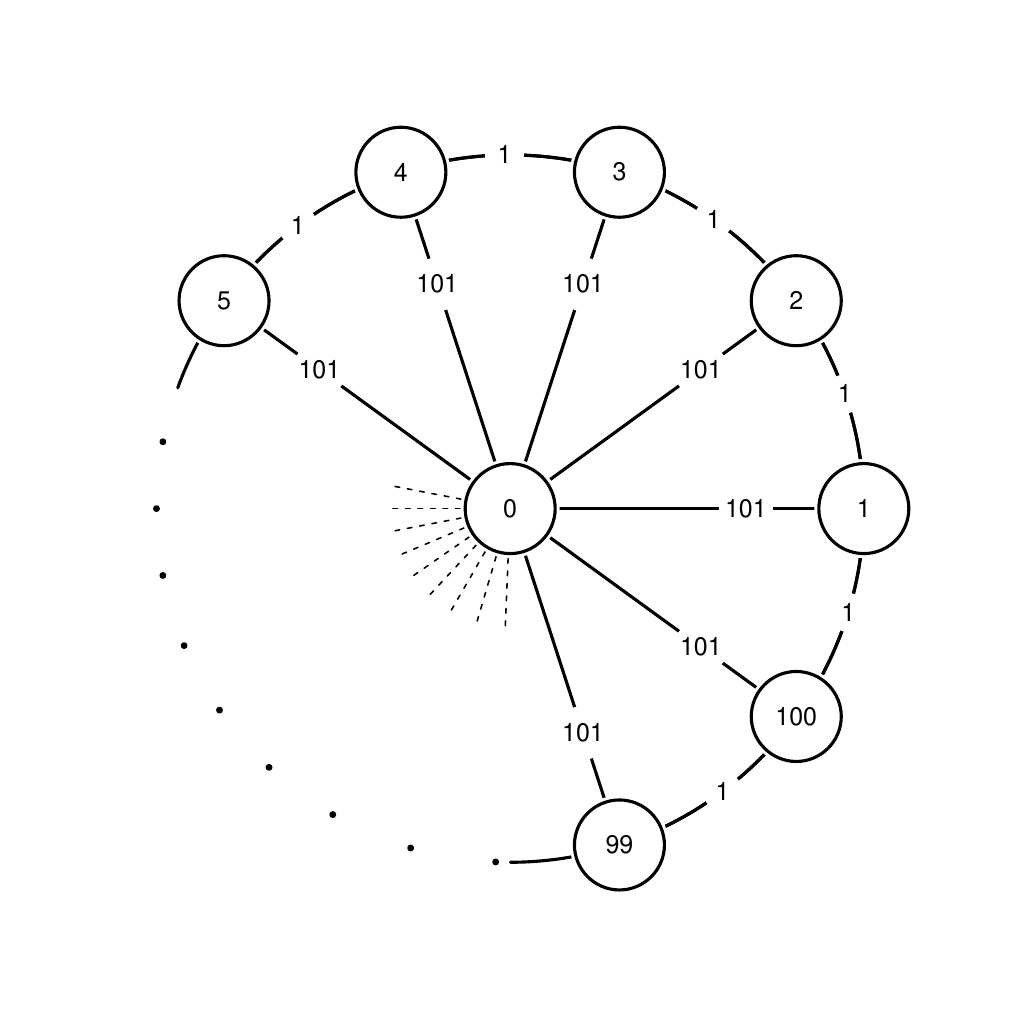}
		\caption{Graph of MinimumSpanningTreeGame}
		\label{minimumspanningtreerajz}
	\end{center}
\end{figure}
\noindent
All the weights on edges adjacent to node $0$ are $101$, all the weights on edges that connect the circle nodes are $1$. The set of players is $N=\{1,2,\dots,100\}$, the characteristic function is $v(S)= $ sum of edge weights of the minimum spanning tree of subgraph $S\cup\{0\}\subseteq G$.
\end{game}

\begin{game}{bankruptcy game}
\label{game6}
Let $A \in \mathbb{R}^+, n \in \mathbb{N}, \ \ l_1,l_2,\dots l_n \in \mathbb{R}^+$ and let game $(N,v)$ be an \mbox{$n$-person} game where the characteristic function $v$ is as follows: for all coalitions $S\subseteq N $ \mbox{$v(S) = \max\{0, A - \sum_{j \not \in S } l_j\}$}. For more details on bankruptcy games, see \cite{jatekelm} and \cite{thomson2015axiomatic}. Let us set $n=100$, $A = 200$ and let the vector of liabilities $l_j$ be set to the same numbers as the $c_i$ numbers in Game 4.
\end{game}

\begin{game}{liability game}
\label{game7}
Liability games have recently been introduced by \cite{csoka2019liability}. Let $A \in \mathbb{R}^+, n \in \mathbb{N}, \ \ l_1,l_2,\dots l_n \in \mathbb{R}^+$ where $A < \sum l_i$, and let game $(N,v)$ be an \mbox{$n$-person} game with characteristic function \[v(S) = \begin{cases} \min\{ A, \sum \limits_{j  \in S \setminus \{0\} } l_j\} & \text{if} \ 0 \in S \\ \max\{ 0, A- \sum \limits_{j  \notin S \cup \{0\} } l_j\} & \text{if} \ 0 \notin S \\\end{cases}\,.\]
According to \cite{csoka2022shapley}, the Shapley value of liability games cannot be calculated in polynomial time (unless P=NP), so the estimation of their Shapley value is particularly interesting. In our example we set $n=100$, $A = 200$ and the vector of liabilities $l_j$ are set to the same numbers as the $c_i$ numbers in Game 4.
\end{game}

\begin{game}{pair game}
\label{game8}
Let $n=100$, $N = N_1 \cup N_2$, $|N_1|=|N_2|=50$ and let $G(N_1, N_2, E)$ be a bipartite graph with a complete matching (to put it simply, each player has a pair). The characteristic function of the game is $v(S)=1/2 \cdot |(i,j) \in S \times S \ \text{such that} \ (i,j) \in E |$, i.e. the number of pairs in the coalition.
\end{game}
\subsection{Analytical results}
\begin{sloppypar}
Algorithm OptK2Shapley practically consists of two main steps: first, it uses some of the available computational power to find an optimal permutation that can be used to generate two correlated samples (Algorithm GreedyK2). After that, it uses the remaining computational power to actually generate these correlated samples and provides an empirical Shapley value as an estimate. The sample size of the first step is $m_1$ and the total sample size that the second step generates is $2m_2$. We set these parameters such that the total complexity of the algorithm is the same as the generation of $m=2m_2+\frac{n^2}{6}m_1$ independent samples.

The standard deviation of simple random sampling is $\frac{\sigma}{\sqrt m}$ where $\sigma$ is the standard deviation of the marginal contributions, and $m$ is the sample size. Algorithm OptK2Shapley generates the sample pairwise, $Y_{1,1,} Y_{2,1},Y_{1,2}, Y_{2,2}, \dots ,Y_{1,m_2}, Y_{2,m_2}$, where $Y_{1,j}$ and $Y_{2,j}$ are identically distributed and not independent but the two-dimensional variables $(Y_{1,1},Y_{2,1}),(Y_{1,2},Y_{2,2}), \dots ,(Y_{1,m_2},Y_{2,m_2})$ are independent and have identical joint distribution.
The variance of the average of such variables can be calculated as a function of the correlation between the coordinates $\rho = \text{cor}(Y_1,Y_2)$ (which is to be minimized by algorithm GreedyK2).
\end{sloppypar}
\begin{multline*}
D^2\left(\frac{ Y_{1,1,}+ Y_{2,1}+Y_{1,2}+ Y_{2,2}+\dots+Y_{1,m_2}+Y_{2,m_2}}{2m_2} \right)=\\
=D^2\left( \frac{\frac{Y_{1,1}+Y_{2,1}}{2} + \frac{Y_{1,2}+Y_{2,2}}{2} + \dots + \frac{Y_{1,m_2}+Y_{2,m_2}}{2} }{m_2} \right)=\\
=\frac{1}{m_2^2}\cdot m_2 \cdot D^2\left( \frac{Y_{1,1}+Y_{2,1}}{2} \right)=\\
=\frac{1}{m_2}\cdot \frac{1}{4}\cdot \left( D^2\left(Y_{1,1}\right)+ D^2\left(Y_{2,1}\right)+2\cdot cov(Y_{1,1},Y_{2,1}) \right)=\\
=\frac{1}{4m_2}\cdot \left(\sigma^2+\sigma^2+2\cdot \sigma \cdot \sigma \cdot \rho \right)=
\frac{1}{4m_2}\cdot\left(  2\sigma^2 \cdot (1+\rho)  \right)=
\frac{\sigma^2}{2m_2}\cdot\left( 1+\rho  \right)\,.
\end{multline*}
Taking its square root, we conclude that the standard deviation of the sample is the standard deviation of the marginal contributions divided by the sample size (like in the case of an independent sample) and multiplied by the factor $\sqrt{(1+\rho)}$. If the correlation $\rho$ is negative, this is an obvious improvement by the price of having a lower sample size $2m_2 = m - \frac{m_1n^2}{6}< m$. Since some of the computational power has been used up to find the transformation that makes the variates negatively correlated, we can generate a smaller sample. The algorithm outperforms simple random sampling (i.e., has lower variance) if and only if
\[\frac{\sigma}{\sqrt{2m_2}}\sqrt{\left( 1+\rho  \right)} \le \frac{\sigma}{\sqrt m} \Longleftrightarrow m\cdot(1+\rho) \le 2m_2\,.\]
In this case, the improvement (the factor by which we can reduce the standard deviation of the simple random sampling) is
\begin{equation}
\label{eq:lambda}
\frac{\sigma_E}{\sigma_R}= \frac{\frac{\sigma}{\sqrt{2m_2}}\sqrt{1+\rho}}{\frac{\sigma}{\sqrt m}}=\sqrt{\frac{m}{2m_2}(1+\rho)}\,.
\end{equation}
As $2m_2 = m - \frac{m_1n^2}{6}$, from Eq. \ref{eq:lambda} we obtain the upper bound 
\begin{equation}
\label{egyenlotlenseg1}
m_1 \le \frac{-6\rho m}{n^2}\,.
\end{equation}
Considering that $\rho$ in most cases is a decreasing function of $m_1$, it can be positive for lower values of $m_1$, in which cases this condition cannot be satisfied at all. 
On the other hand, if we fix an $m_1$ that is high enough to get a negative $\rho$ and let the sample size go to infinity: if $m\rightarrow \infty$, then obviously $\frac{m}{2m_2}\rightarrow 1$, which means that $\sigma_E/\sigma_R \rightarrow \sqrt{1+\rho}$. Algorithm OptK2Shapley asymptotically grants this ratio of improvement.

Now, the only thing that is still missing to determine the minimum value of $\sigma_E/\sigma_R$ (Eq. \ref{eq:lambda}) is to know how low correlation $\rho$ can be achieved. Rearranging the terms in Eq. \ref{egyenlotlenseg1}, we obtain that the algorithm outperforms simple random sampling if and only if
\begin{equation}
\label{egyenlotlenseg2}
\rho \le -\frac{n^2\cdot m_1}{6m}\,.
\end{equation}
Let us introduce the notation \[\bar{\rho} = \min \{\rho(Y,Y^t) \ | \ t:S_n \leftrightarrow S_n \text{transformation such that} \ t^2 = id\},\]
where $Y$ is a random variable of marginal contributions, and $Y^t$ is the transformed variable according to the ergodic sampling algorithm. So $\bar{\rho}$ is the theoretical optimum of the possible correlations the algorithm tries to minimize, which depends on the game and the player in question. By definition, $\mathbb{P}(\bar{\rho} \le \rho) = 1$ and $\bar{\rho}$ is independent of the sample size or the realizations of the MC algorithm, or anything else in the algorithm, it is characterized by the game, and in certain cases, it can be calculated or estimated analytically. For example, in the case of the symmetric voting game\footnote{It is $\bar{\rho} = -\frac{1/n}{1-1/n}$ see the comments after Algorithm ShapleyK3 in Section 4.} $\bar{\rho} = -1/(n-1)$, and the optimal transformation $t$ (which is not unique) is to swap the pivot player and any other player. In this case, $\bar{\rho}$ is very close to zero, so we do not expect the algorithm to perform well, but at least it does not increase the estimation error. Less trivial, but in the case of the minimum spanning tree game, we have that $\bar{\rho} \le -98.51\%$ (transformation $t$ is the ``mirror image'' of the original order), so we expect the algorithm to perform well in the case of this game. 

\subsection{Simulation results}
\begin{figure}
	\begin{center}
		\includegraphics[scale = 0.65]{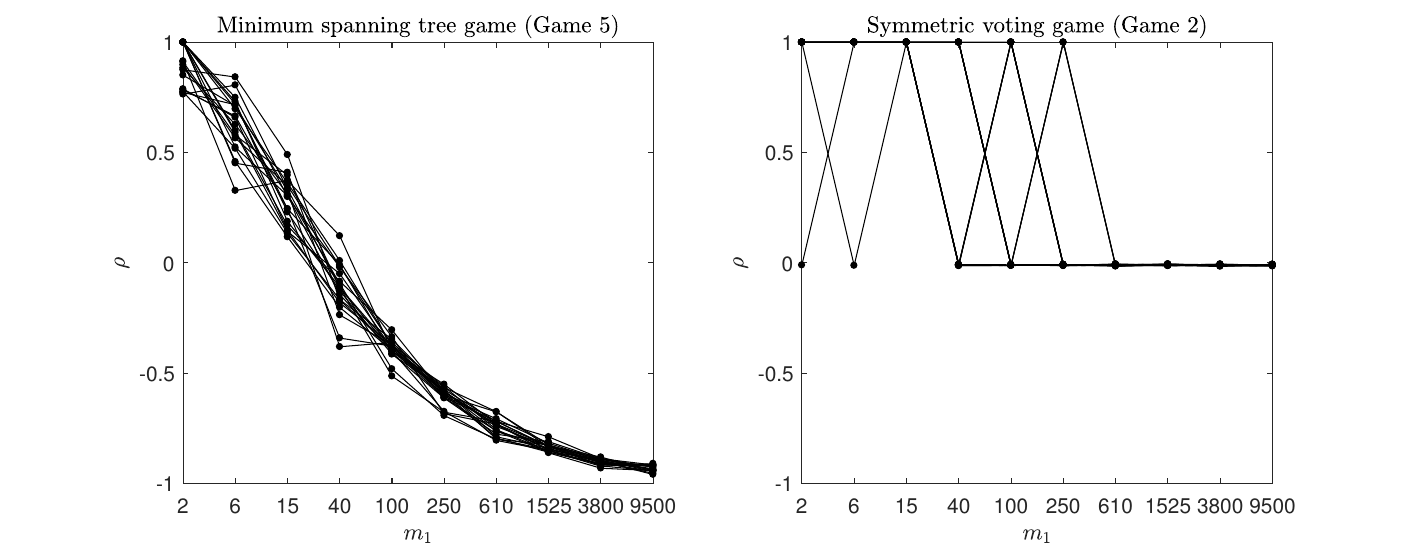}
		\caption{The figures show the correlation $\rho$ between the samples as a function of the sample size $m$ based on 20 random samples. In the case of the minimum spanning tree game the correlation decreases and converges to a ``highly'' negative value. In the case of the voting game there is a random critical value of $m_1$ where the correlation falls to zero and then stays there.}
		\label{Game5cor}
	\end{center}
\end{figure}

\begin{sloppypar}Figure \ref{Game5cor} shows a simulation of correlation $\rho$ by algorithm GreedyK2 (\mbox{Algorithm \ref{alg:greedy}}) for different values of $m_1$ in case of the minimum spanning tree game (Game \ref{game5}) and the symmetric voting game (Game \ref{game2}).
\end{sloppypar}
In the case of the minimum spanning tree game, we can see that the correlation is quite stable even for moderate values of $m_1$. If we set $m_1 = 100$, we can expect a correlation approximately $-0.5$, for $m_1 = 1000$ it is about $-0.8$ with low standard deviation. In the case of the symmetric voting game, we can see a jump from 1 to $-1/99$. If $m_1$ is high enough, algorithm GreedyK2 learns to swap the 51st player with any other random player, which gives practically no correlation between the samples. In this case, $\rho$ equals its theoretical minimum $\bar{\rho}$, and no further improvement is possible. These observations suggests that the limit $\rho^*=\lim_{m_1 \rightarrow \infty} \rho(m_1)$ exists with probability 1, and is deterministic (the variance of the correlation seems to vanish as the sample size $m_1$ increases). It is reasonable to believe that $\rho^*= \bar{\rho}$ or very close. 

\begin{sloppypar}
Table \ref{tabRes} shows the results of a numeric simulation. To analyze the mechanism of the algorithm and the effects of the parameters, we have run the algorithm OptK2Shapley for different parameter settings $1000$ times, and we have estimated the standard deviation of the empirical Shapley values as a percentage of the standard deviation of simple random sampling\footnote{A Matlab code that implements these algorithms is available at \url{https://www.mathworks.com/matlabcentral/fileexchange/71822-estimation-of-the-shapley-value-by-ergodic-sampling}.}. The first four columns show the parameters and the fourth one shows simulated correlation by algorithm Algorithm GreedyK2. Finally, the last two columns show the theoretical and the empirical value of the standard deviation of ergodic sampling as a percentage of simple random sampling by \mbox{formula (\ref{eq:lambda})} and by the numeric simulation. 
\end{sloppypar}

\begin{table}
\centering
\begin{threeparttable}
\small
{
\renewcommand{\arraystretch}{0.9}
	\begin{tabular}{crrrrcc}
		\toprule
		\multicolumn{1}{c}{\textbf{Game}} & \multicolumn{1}{c}{$\mathbf{m}$} & \centering $\mathbf{m_1}$ & \multicolumn{1}{c}{$\mathbf{m_2}$} & \multicolumn{1}{c}{$\boldsymbol\rho$} & \multicolumn{1}{c}{$\sqrt{\frac{m}{2m_2}(1+\rho)}$} & \multicolumn{1}{c}{$\mathbf{\hat{\sigma}_E/\sigma_R}$} \\
		\hline
		\multirow{5}{*}{$\substack{\text{Game 1}\\\\\text{Non-symmetric}\\\text{voting game}\\(51,1)}$} & 100k&100&28\,325&0.0546&1.36&1.34\\
			&600k&100&278\,325&0.0546&1.06&0.99\\
			&4000k&100&1\,978\,325&0.0546&1.03&0.96\\
			&600k&1000&83\,250&0.0180&1.94&1.95\\
			&4000k&1000&1\,783\,250&0.0180&1.08&1.03\\
			\hline
		\multirow{5}{*}{$\substack{\text{Game 2}\\\\\text{Symmetric}\\\text{voting game}\\(100,100)}$}&1000k&500&83\,333&-0.0069&2.44&2.39\\
			&2000k&50&958\,334&1.0000&1.44&1.43\\
			&10\,000k&50&4\,958\,334&-0.0090&0.99&0.98\\
			&2\,000k&1000&166\,667&-0.0101&2.43&2.43\\
			&10\,000k&1000&4\,166\,667&-0.0101&1.08&1.12\\
			\hline
		\multirow{5}{*}{$\substack{\text{Game 3}\\\\\text{Shoes game}\\(100,100)}$}&400k&100&116\,667&-0.3017&1.09&1.07\\
			&3000k&100&1\,416\,667&-0.3017&0.85&0.79\\
			&10\,000k&100&4\,916\,667&-0.3017&0.84&0.81\\
			&3000k&1500&250\,000&-0.7016&1.33&1.30\\
			&10\,000k&1500&3\,750\,000&-0.7016&0.63&0.64\\
			\hline
		\multirow{5}{*}{$\substack{\text{Game 4}\\\\\text{Airport game}\\(100,100)}$}&200k&50&58\,333&-0.0071&1.30&1.28\\
			&2000k&50&958\,333&-0.0071&1.02&1.02\\
			&7000k&50&3\,458\,333&-0.0071&1.00&0.97\\
			&2000k&500&583\,333&-0.0109&1.30&1.30\\
			&7000k&500&3\,083\,333&-0.0109&1.06&1.04\\
			\hline
		\multirow{5}{*}{$\substack{\text{Game 5}\\\\\text{Minimum spanning}\\\text{tree game}\\(100,1)}$}&250k&100&41\,667&-0.4266&1.31&1.26\\
			&1700k&100&766\,667&-0.4266&0.79&0.85\\
			&4000k&100&1\,916\,667&-0.4266&0.77&0.83\\
			&1700k&1000&16\,667&-0.8108&3.10&3.19\\
			&4000k&1000&1\,166\,667&-0.8108&0.56&0.57\\
			\hline
		\multirow{5}{*}{$\substack{\text{Game 6}\\\\\text{Bankruptcy game}\\(100,100)}$}&200k&100&16\,667&-0.5291&1.68&1.76\\
			&1000k&100&416\,667&-0.5291&0.75&0.74\\
			&10\,000k&100&4\,916\,667&-0.5291&0.69&0.70\\
			&1000k&500&83\,333&-0.5606&1.62&1.62\\
			&10\,000k&500&4\,583\,333&-0.5606&0.69&0.70\\
			\hline
		\multirow{5}{*}{$\substack{\text{Game 7}\\\\\text{Liability game}\\(101,0)}$}&200k&100&16\,667&-0.6888&1.36&1.36\\
			&1000k&100&416\,667&-0.6888&0.61&0.70\\
			&10\,000k&100&4\,916\,667&-0.6888&0.56&0.61\\
			&1000k&500&83\,333&-0.9454&0.57&0.56\\
			&10\,000k&500&4\,583\,334&-0.9454&0.24&0.26\\
			\hline
		\multirow{5}{*}{$\substack{\text{Game 8}\\\\\text{Pair game}\\(100,100)}$}&250k&100&41 667&-0.2954&1.45&1.52\\
			&2000k&100&916\,667&-0.2954&0.87&0.79\\
			&5000k&100&2\,416\,667&-0.2954&0.85&0.88\\
			&2000k&1000&166\,667&-0.7264&1.28&1.26\\
			&5000k&1000&1\,666\,667&-0.7264&0.64&0.67\\
			\bottomrule
	\end{tabular}
}
\end{threeparttable}
\normalsize

\caption{The table shows the estimation error of the algorithm for various parameter settings. The numbers in the parentheses are the number of players in the game and the player's id whose Shapley value we calculate, respectively.}
\label{tabRes}
\end{table}

Based on the table, we can make the following conclusions:
\begin{itemize}
\item The theoretical end empirical values of the improvement in the estimation error match aside from a small noise in the simulation.
\item There is a nontrivial trade-off between parameters $m_1$ and $m_2$. If $m_1$ is large, we can find a better transformation with a stronger negative correlation by the price of having a lower value for $m_2$. If $m_2$ is very small, finding a negative correlation is not worth it, so increasing $m_1$ can have a positive and a negative effect as well. 
\item In exceptional cases, if $m_1$ is too small, the correlation can be unstable. For example, in the second and third rows of \mbox{Game 2} $m_1$ is the same, but the correlation is $1$ randomly in the first case (with insufficient information, the algorithm learns nothing, it merely duplicates an independent sample ($\rho = 1$), which multiplies the standard deviation by $\sqrt 2$) and about $0$ in the second one.
\item After a certain threshold (if $\rho$ is close to $\rho^*$), increasing $m_1$ does not grant a significantly stronger correlation.
\item Even in the worst cases, $\rho^*$ and $\bar{\rho}$ are very close to zero, and in five of eight cases, it is strongly negative.\footnote{For each game there is at least one row in the table where the value in the last column is at most one and it some cases, it is much smaller.} Practically, three test cases resist the algorithm, but the estimation error is significantly decreased in five cases.
\end{itemize}

\begin{figure}
	\begin{center}
		\includegraphics[scale = 0.65]{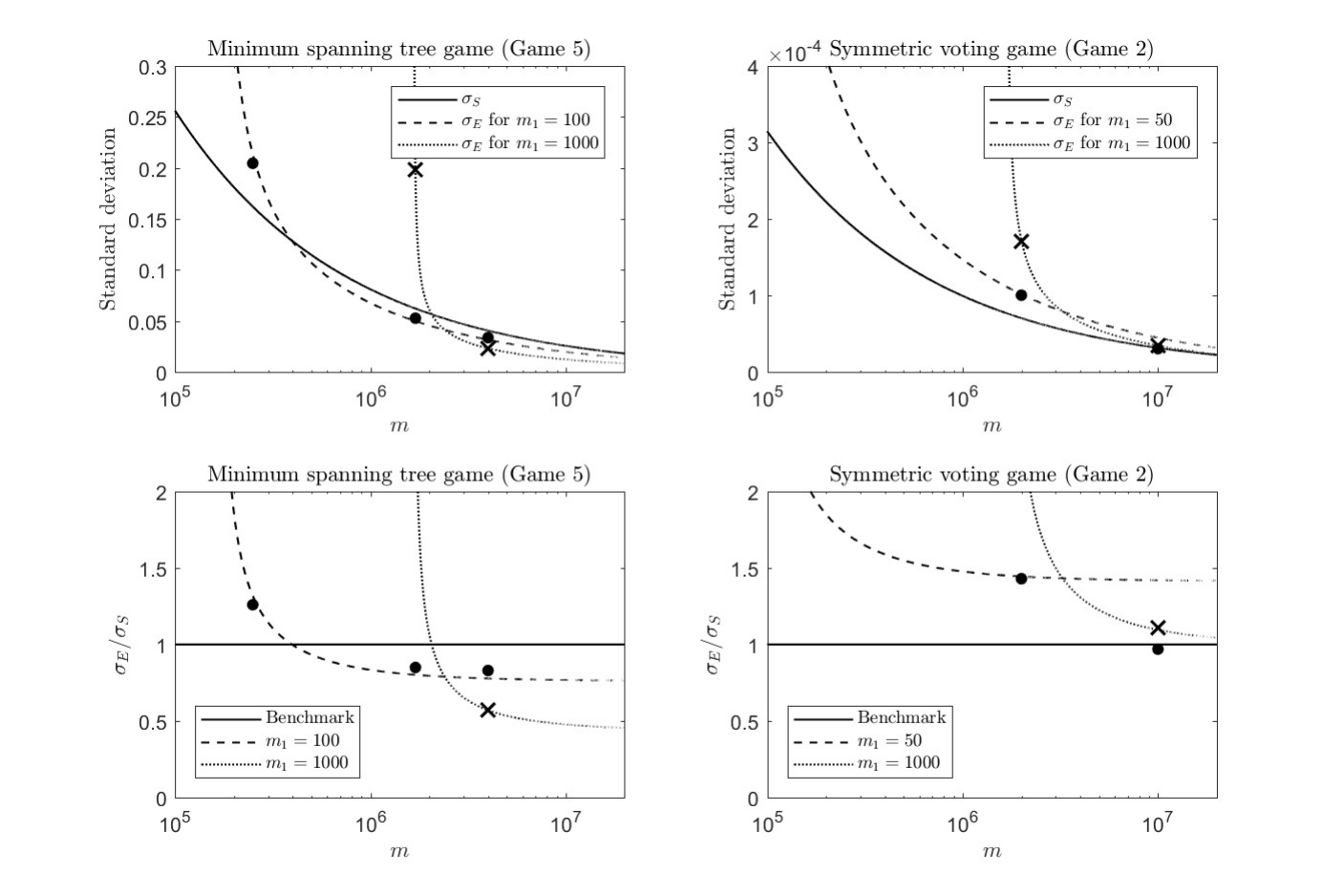}
		\caption{Figures show the performance of the algorithm as a function of the sample size for the minimum spanning tree game and the symmetric voting game. The ergodic sampling outperforms the simple random sampling if the dashed and dotted lines intersect the solid line. This is true for the minimum spanning tree game but not in the case of the voting game. Markers represent the computational result from Table \ref{tabRes}, and they match the theoretical results.}
		\label{Game5Graph}
	\end{center}
\end{figure}

In Figure \ref{Game5Graph}, the results are even more transparent. The top-left panel shows the standard deviation of the estimated Shapley value of the minimum spanning tree game for simple random sampling as a benchmark and for OptK2Shapley with $m_1=100$ (dashed line) and $m_1 = 1000$ (dotted line) as a function of the sample size of the benchmark ($m$). Obviously, the standard deviation of OptK2Shapley only exists if $m \ge m_1\frac{n^2}{6}$, at the point where $m_2 = 0$ there is a ``singularity''. As the correlation is negative in both cases, there is a critical value of $m$ above which it is worth running algorithm OptK2Shapley instead of simple random sampling. This is the point where $m=\frac{n^2m_1}{-6\rho}$. There is also a point that divides the values of $m$ where $m_1=100$ is better than $m_1=1000$. This critical value is the point where $\frac{m_2(m_1=100)}{m_2(m_1=1000)}=\frac{1+\rho(m_1=100)}{1+\rho(m_1=1000)}$. We can see that if $m\rightarrow \infty$, it is worth choosing the higher $m_1$ value if it improves the value of $\rho$ further as this curve has a lower limit in infinity. In the bottom-left panel, we can see the standard deviation of ergodic sampling divided by the standard deviation of simple random sampling, and we also indicated on the graph the points corresponding to the results of simulations in \mbox{Table \ref{tabRes}}. When the lines cross the horizontal solid line, the algorithm performs better than simple random sampling. We can see the same graphs for the minimum spanning tree game on the right side, for $m_1 = 50$ and $m_1 = 1000$. Though the algorithm performs moderately in both cases, there is a critical difference between them. When choosing $m_1 = 50$ (dashed line), the algorithm does not have enough information to learn the transformation that swaps the player in position $51$ with someone else. This happens because there is no observation where Player 100 is in position $51$ in the arriving order. Nevertheless, this is very unlikely if we have a thousand observations.

From this analysis, we can conclude that the relationship between the algorithm's parameters and their effects on the results is very well understood. Now we have a good guess on how to set the parameters. In the case of the symmetric voting game, the algorithm has no chance to learn the optimal transformation if there is no observation when player $i$ is the pivot player. This suggests that (as a minimum requirement) for each $j \in \{1,2,\dots, n\}$ the sample should contain observations where player $i$ is in position $j$. This is obviously impossible if $m_1<n$, but its probability converges to 1 quickly even if the sample size is linear in $n$. For example if $m_1 = c\cdot n$ then the probability of a sample of $m_1$ independent observations not containing cases where player $i$ is the pivot player in a voting game is \[\left(1-\frac{1}{n}\right)^{m_1}=\left(1-\frac{1}{n}\right)^{cn}\approx e^{-c}.\]
This is about $5\%$ for $c = 3$ and negligibly small for $c = 10$. The numerical experiments in Table \ref{tabRes} also suggest that the algorithm is unlikely to improve the correlation a lot after $m_1 \approx 10n$.

\section{Conclusions and further research}
\label{sec7}
Though the Shapley value is a fundamental solution concept of cooperative games, its exact calculation has proven impossible for many game classes (unless \mbox{P = NP}). However, in the case of games with efficiently computable characteristic function (i.e., polynomial in the number of players), the Shapley value can be approximated using various \mbox{Monte-Carlo} methods. 

The main contribution of this paper to the literature is that \mbox{Monte-Carlo} simulation does not necessarily give the best result if we generate an independent sample. A negative correlation between subsamples reduces the variance of the mean of the entire sample. This idea in statistics is called the method of antithetic variates, and its use to estimate the Shapley value is a novelty in the literature. As we have only presented the main idea for now and proven that it works, there are many possible extensions and further research questions. First, a more sophisticated and maybe faster algorithm to find a suitable ergodic transformation could be useful. It is also unsolved how to find a higher-order transformation than two.

To put the question into a wider perspective, three more important questions would be wonderful to answer:
\begin{itemize}
\item How can we measure the goodness of a variance reduction method in the case of the Shapley value? (Current results in the literature can only provide illustrative examples, not statistical results, as there is no such thing as a ``typical TU game".)
\item How to characterize the game classes for which a given MC method works well? This may not even be possible as there can be a strong dependency on the game parameters, not just on the type of the game.
\item Which methods can be combined to get an even better method?
\end{itemize}

\newpage






\bibliographystyle{plainnat}
\bibliography{references2}




\end{document}